\documentstyle[12pt]{article}
\begin{document}
\title{Redistribution of the Hole Spectral Weight due to Long-Range Spin 
Correlations in the Three-Band Hubbard Model} 
\author{A.F.Barabanov$^a$, L.A.Maksimov$^b$, E.\v Z\c asinas$^a$, 
O.V.Urazaev$^a$\\ 
{\small $^a$Institute for High Pressure Physics, Troitsk, Moscow 
region, 142092, Russia.}\\
{\small $^b$ Russian Research Centre, 
Kurchatov Institute, Kurchatov sq.46, Moscow, 123182, Russia.}} 
\date{}
\maketitle \begin{abstract}
In the framework of the three-band model for $CuO_{2}$ plane in high-temperature 
superconductors the spectrum  of the spin-polaron hole exitation is investigated.
The problem is treated taking into account the coupling of a local polaron with the
antiferromagnetic spin wave with $\bf Q=(\pi,\pi)$.This leads to the essential
changes of the lowest polaron band $\epsilon_{1}(\bf k)$ and the strong 
redistribution of the bare electron filling.

PACS number(s): 74.72.-h. 75.10-b
\end{abstract} 
In order to understand the nature of high-temperature superconductors it is
important to describe properly the motion of a hole in the $CuO_2$ plane 
\cite{dagotto,brenig}. This motion takes place on the antiferromagnetic (AFM) spin
background of copper spins and it must be treated as a correlated motion of
a hole coupled to spin excitations (a spin polaron). Usually the spin
polaron is studied within the frameworks of the t-J model \cite{dagotto} and
the three-band Hubbard model \cite{emery,zh,bar}.
In the previous works the spin polaron problem was studied mainly in the 
approximation  of a small radius polaron ( an analogous of a Zang -Rise
polaron). In the present work, for the first time, the long range order
of the AFM background is taken into account by introducing an additional new
spin polaron of an infinite radius -a bound state of a charge exitation
and a spin wave with $\bf Q=(\pi,\pi)$.We shall show that the introduction 
of such a polaron leads  to the essential decrease of lowest band
filling by bare holes.As a result the area of the Fermi surface
strongly increases and its form becomes more complex.

The effective Hamiltonian of the three-band Hubbard model of the 
$CuO_2$ plane has the following form in conventional notations 
\cite{emery,matsukava,bar}:
\begin{eqnarray}\label{hamilton}
\hat H=\hat T+\hat h+\hat J,\; \hskip 20 pt
\hat T=\tau\sum_
{
{{{\bf R},{\bf a}_1,{\bf a}_2,}\atop
{{\sigma}_{1},{\sigma}_{2}}}
}
{X}_{\bf R}^{{\sigma}_1{\sigma}_2}
{c}^{+}_{{\bf R}+{\bf a}_2,{\sigma}_2}
{c}^{\phantom{+}}_{{\bf R}+{\bf a}_1,{\sigma}_1}\;\\
\hat h =-h\sum_{{\bf R,a,b},\sigma} {c}^{+}_{{\bf R+a},\sigma}
{c}^{\phantom{+}}_{{\bf R+a+b},\sigma}\; ,\hskip 30 pt
\hat J={J\over 2}\sum_{\bf R,g} {\bf \hat S}_{\bf R} {\bf \hat
S}_{\bf R+g}\; . \nonumber
\end{eqnarray}

Here the $CuO_2$ plane is described by the square sublattice with
lattice constant $g$ and two $O$ sites per a $Cu$ unit cell; ${\bf R}$
--- the vectors of $Cu$ sites, ${\bf R+a}$ --- are four vectors of $O$
sites nearest to $Cu$ site ${\bf R}$, ${\bf a}=\pm {\bf a}_x,\pm {\bf
a}_y$, ${\bf a}_x=g(\frac{1}{2},0)$, ${\bf a}_y=g(0,\frac{1}{2})$. 
We assume $g=1$. In~(\ref{hamilton}) we take the notations: {\bf b} 
are the nearest neighbor (NN) vectors for the oxygen sublattice; 
${\bf g}=2{\bf a}$ and ${\bf d}=2{\bf b}$ are the first- and second- 
NNs for $Cu$ sublattice; the operators ${c}^{+}_{\sigma}$ and 
${X}^{\sigma 0}$ create a hole with spin $S=\frac {1}{2}$ and spin 
projection $\frac {\sigma}{2}$ $(\sigma =\pm 1)$ at the $O$ and $Cu$ 
sites respectively and ${X}_{\bf R}^{{\sigma}_{1}{\sigma}_{2}}$ are 
the Hubbard projection operators which are convenient for excluding 
doubly occupied $Cu$ sites.

The first term $\hat T$ in (\ref{hamilton}) describes the effective hole
hopping with amplitude  from $O$ to $O$ sites
through the intervening $Cu$ sites. The term $\hat J$
correspond to AFM interaction between $Cu$ sites.
$\hat h$ represents the direct $O$--$O$ NN hopping. 

Let us discuss the hole excitations with spin $S=\frac{1}{2}$ and the
spin projection $\frac{\sigma}{2}$. We shall restrict ourselves to a
finite number of site operators $A_{{\bf R},j}, (1 \leq j \leq 12)$,
in each unit cell ${\bf R}$.

In order to treat the hole excitations in the framework of the 
spin-polaron concept we introduce for each unit cell $\bf R$ six site 
operators which describe the local polaron of small radius.  
\begin{eqnarray}
A^+_{{\bf R},\sigma,1(2)}=c^+_{{\bf R}+{\bf a}_x({\bf a}_y) 
,\sigma}\; ,\hskip 20 pt A^+_{{\bf R},\sigma,3(4)}
=\sigma\sum_{\gamma =\pm 1}\gamma X_{\bf 
R}^{\overline{\gamma}\:\overline{\sigma}} c^{+}_{{\bf R}+{\bf 
a}_x,({\bf a}_y),\:{\gamma}}\;\label{oper16} \nonumber\\ 
A^+_{{\bf R},\sigma,5(6)}=\sigma\sum_{\gamma =\pm 1}\gamma X_{{\bf 
R}+{\bf g}_x,({\bf 
g}_y)}^{\overline{\gamma}\:\overline{\sigma}}c^{+}_{{\bf R}+{\bf 
a}_x,({\bf a}_y),\:{\gamma}}\;,\hskip 70 pt  \overline{\sigma}=-\sigma
\end{eqnarray} 
\begin{eqnarray}
A^+_{{\bf k},\sigma,j}={1\over{\sqrt{N}}}
\sum_{\bf R} {\rm e}^{i{\bf kR}} A_{{\bf R},\sigma,j}\; .\nonumber
\end{eqnarray} 

$A^+_{{\bf k},\sigma,j}$ are the Fourier transforms of
$A^+_{{\bf R},\sigma,j}$.

Let us mind, that in our previous investigations \cite{key} this 
basis of local spin-polaron operators lead to the proper description 
of the experimentally observed important features of the $CuO_2$ 
plane hole spectrum: extended saddle point and isotropic band bottom. 

In this paper we want to investigate the role of the delocalized spin 
polarons which correspond to the coupling of the local polarons to 
AFM spin wave with momentum ${\bf Q}=(\pi,\pi)$, so 
called Q-polarons usually, if the spin subsystem is found in the state with
the AFM long range order then the average value of the amplitude
 $\langle{\bf S}_{\bf Q}\rangle$
 of the spin wave with ${\bf q}={\bf Q}$ (the Q-wave) has 
the macroscopic large value and has the properties analogous to the 
amplitude of a Bose particle with the zero momentum in the superfluid 
Bose-gas. As a result for many problems this amplitude can be treated 
as a $c-$number~\cite{schrieffer}. Then the coupling of the Q-wave 
to local electron states does not represent new states but leads, as 
mentioned above, to the mixing of the states with the momenta {\bf k} 
and {\bf k+Q}. This treatment is usually based on the widely used 
Neel type state of the spin subsystem with two sublattices. 

But for the S=1/2 spin system the quantum fluctuations are very 
important and they lead to the spherically symmetric homogeneous Neel 
state at any small but finite temperature. In this background even at 
$T=0$ the average value $\langle{\bf S}_{\bf Q}\rangle=0$ and the 
mentioned above simple approach for the hybridization of {\bf k} and 
{\bf k+Q} states fails.  In the homogeneous Neel state only 
$\langle{\bf S}_{\bf Q}{\bf S}_{\bf Q}\rangle$ can be treated as a 
macroscopic value. Then the coupling of a local polaron states to 
${\bf S}_{\bf Q}$ corresponds to a new delocalized states. In order 
to take these states into account we introduce the additional six 
operators based on the basis of (\ref{oper16}):  
\begin{eqnarray}\label{oper712}
A^{+}_{{\bf R},\sigma,j}=\sigma\sum_{\gamma =\pm 1}\gamma 
Q_{\bf R}^{\overline{\gamma}\:\overline{\sigma}} 
A^{+}_{{\bf R},\gamma,i},\hskip 56 pt 
A^{+}_{{\bf k},\sigma,\gamma}=\sigma\sum_{\gamma}\gamma
A^{+}_{\bf{k+Q},\gamma,i}
\bf S^{\overline{\gamma}\:\overline{\sigma}}_{\bf Q},\\
Q_{\bf R}^{\overline{\gamma}\:\overline{\sigma}}\equiv{\rm e}^{i{\bf QR}}
S^{\overline{\gamma}\:\overline{\sigma}}_{\bf Q}=
N^{-1}\sum_{{\bf R_1}}e^{i{\bf Q(R+R_1)}}
X _{\bf R_1}^{\overline{\gamma}\:\overline{\sigma}},\hskip 10 pt 
j=i+6,\; i=(1\div6) \;.\nonumber
\end{eqnarray}

In order to determine the spin polaron spectrum $\varepsilon_i({\bf 
k})$ of 12 quasiparticle bands we use the two time retarded matrix 
Green's functions $G_{i,j}(t,{\bf k})$ for the operators $A_{{\bf 
k},\sigma,i}$:  
\begin{equation} \label{gr}
G_{i,j}(t,{\bf k})\equiv \left\langle 
A^{\phantom{+}}_{{\bf k},i}(t)\big\vert A^+_{{\bf k},j}(0) 
\right\rangle =-i\Theta(t) \left\langle\left\lbrace 
A^{\phantom{+}}_{{\bf k},i}(t)\; ,A^+_{{\bf k},j}(0)
\right\rbrace\right\rangle\; ,
\end{equation}

We solve the system of the equations of motion for $G_{i,j}({\omega},{\bf 
k})$ by using the standard Mori-Zwanzig projection technique and 
restricting ourselves to the above chosen basis of operators 
$\left\lbrace A_{{\bf k},\sigma,i}\right\rbrace$ 
(\ref{oper16},\ref{oper712}).  Then the Green's functions and the 
spectrum are determined from the equations:  
\begin{equation}
\left(\omega -DK^{-1}\right)G=K\: ,\hskip 10 pt
det\vert K\varepsilon({\bf k})-D\vert=0.  
\end{equation}
\begin{equation}
{D}_{i,j}({\bf k})=
\left\langle\left\lbrace B^{\phantom{+}}_{{\bf k},i}\; ,
A^{+}_{{\bf k},l}\right\rbrace\right\rangle,\hskip 10 pt
K_{i,j}=\left\langle\left\lbrace
A^{\phantom{+}}_{{\bf k},i}\; ,A^+_{{\bf k},j}
\right\rbrace\right\rangle\; ,\hskip 10pt
B^{\phantom{+}}_{{\bf k},i}=
\left\lbrack A^{\phantom{+}}_{{\bf k},i}\; ,H\right\rbrack\; .
\end{equation}

The matrix elements of $D$ and $K$ are expressed both through
short-range spin-correlation functions of $Cu$ subsystem and the 
long-range order correlation function $\langle{\bf S}_{\bf 
Q}{\bf S}_{\bf Q}\rangle$.  The $Cu$ spin subsystem is described by 
the $S={1\over 2}$ Heisenberg model at $T=0$. For the value of the 
correlation functions we use the results of \cite{berezovsky} where 
this model was treated in the framework of spherically symmetrical 
Green's functions theory.
It is important that we take into account that $\langle{\ bf S}_{\bf 
Q}{\bf S}_{\bf Q}\rangle$ is a macroscopic value and is equal to a square 
of effective sublattice magnetization:  
\begin{equation}
\langle{\bf S}_{\bf Q}{\bf S}_{\bf Q}\rangle=
\lim\limits_{{\bf R_1}\to\infty}\big\vert \langle {\bf S}_{\bf R}{\bf
S}_{{\bf R+R_1}}\rangle\big\vert=M^2
\end{equation}

Note that the dependence of the spin polaron excitations spectrum on 
the long-range correlation function $\langle{\bf S}_{\bf 
Q}{\bf S}_{\bf Q}\rangle$ appears only due to treatment of the 
Q-polaron states (\ref{oper712}). Below we take the following 
numerical values of the spin-correlation functions:  $\langle {\bf 
S}_{\bf R}{\bf S}_{{\bf R+g}}\rangle=-0.3521$, $\langle {\bf S}_{\bf 
R}{\bf S}_{{\bf R+d}}\rangle=0.229$, $\langle {\bf S}_{\bf R}{\bf 
S }_{{\bf R}+2{\bf g}}\rangle=0.2$, $M^2=0.0914$. 
The matrix elements were calculated in the low hole doping limit, 
$n<<1$, $n$ -is the total number of oxygen holes per unit cell.
The detailed expressions of matrices $D$ and $K$ will 
be published elsewhere. 

As a result the Green's functions have a form
\begin{equation}
G_{i,j}(\omega,{\bf k})=
\sum_{l=1}^{12}\frac{Z^{(l)}_{(i,j)}({\bf k})}{\omega 
-\varepsilon_l({\bf k})}.  \end{equation} 

In particular the value of 
$Z_h({\bf k})=Z^{(1)}_{(1,1)}({\bf k})+Z^{(1)}_{(2,2)}({\bf k})$ 
corresponds to the number of bare oxygen holes with the fixed spin 
$\sigma$ and the momentum {\bf k} in the state $\vert {\bf 
k},\sigma\rangle$ of the lowest quasiparticle band $\varepsilon_1({\bf k})$. 
Let us mind that residue $Z_{(i,j)}({\bf k})$ satisfy the sum rule
$\sum_{s}Z^{(s)}_{(i,i)}(\bf k)=1,i=1,2$.This means that in this model 
the Luttinger theorem is not fulfilled and the maximum number of holes
per cell is equal four despite the presents of twelve bands.

The results of the most interesting two lowest bands 
$\varepsilon_1({\bf k})$ and $\varepsilon_2({\bf k})$ for realistic 
values  of model parameters $J_1=0.2\tau$, $h=0.4\tau$ (below all 
energetical parameters are expressed in units $\tau$) are presented 
in Fig.1(a). In the same figure the spectrum of the lowest of six bands
$\underline{\varepsilon}_1({\bf k})$, calculated in the approximation 
of six operators (2)  is shown. In 
Fig.1(b) the spectrum $\varepsilon_1({\bf k})$ is presented by 
the equal-energy lines $\varepsilon_1({\bf k})={\rm const}$.
Let us mind that the Q-polarons may lead to a rather complex
form of the Fermi surface.This circumstance can give a nontrivial 
behaviour of the Hall effect on doping, and even cause the inversion 
of the Hall constant if the Fermi energy is close 
to $\varepsilon_{1}(\bf k)=-4.5$.

As it is seen from Fig.1(a), the inclusion of the Q-wave 
qualitatively leads to the decoupling of the lowest band of the local 
small polaron excitations and $\varepsilon_1({\bf k})$ is 
close to $\underline{\varepsilon}_1({\bf k})$. This means that the 
main features of the lowest band excitations previously calculated in 
the local polaron approximation~\cite{key} are preserved. 

The importance of the treatment of the Q-wave polarons may be seen if 
we discuss the filling of the lowest band by bare holes. In  
Fig.1(c) the filling of the ${\varepsilon}_1({\bf k})$ and 
$\underline{\varepsilon}_1({\bf k})$ are shown, i.e., the residues
$Z_h({\bf k})$ and $\underline{Z}_h({\bf k})$ (the underlined values 
correspond to local polaron approximation).
One can see that the introduction of the Q-polarons leads to the 
essential decrease of hole filling along the lines $X$-$M$
and $M$-$N$.This redistribution of the bare hole spectrum weight
explaines the results of photoemission experiments when the``flat band 
region'' is observed along the direction $X$-$\Gamma$
and is invisible along the line $X$-$M$.
Fig.1(d) demonstrates that  the local polaron concept(six operators (2))
leads to the strong decrease of the filling under the Fermi surface.
This means the violation of the Luttinger theorem approximately in four times
(the analogous effect was found in \cite{hayn}).But the inclusion of
a Q-polaron states (3) leads to the addition essential reduction
of filling, approximately in $1.5$ times.The maximum $\varepsilon_{1}(\bf k)$
-band filling is equal to $n=0.22$, and the smallness of this value
justifies our low density approximation.

Formula (3) points out that the Q-polaron
$A^{+}_{\bf k,\sigma,7(8)}$ contains a bare hole state
$c^{+}_{\bf k+\bf Q,\sigma}$. It means that the residual $Z_{\bf Q}$
of the corresponding Green's function $G_{\bf Q}(\bf k)=G_{7,7}+G_{8,8}$
are responsible for ``shadow band'' effect \cite{kam}.

We are to note that Q-polaron scenario reproduces the essential decrease
of the lowest band width with the decrease of the AFM constant J.
Usually such an effect is obtained only in the self-consistent Born
approximation \cite{mar_hor}.

In conclusion we want to mention that the described above properties
of the spin polaron are
mainly preserved if we suppose that the spin subsystem has no long range
order, but the spin correlation length $L$ is large. 
Then, in the matrix elements of $D$ and $K$ we must replace 
the long-range correlation function $\langle{S}_{\bf Q}{S}_{\bf 
Q}\rangle$ by the following expression:
\begin{equation}
\sum_{\vert{\bf q}\vert<\delta q}
\langle{S}_{\bf Q+q}{S}_{\bf Q+q}\rangle \hskip 20pt 
\delta q \approx 1/L,\; L>>1.
\end{equation}

ACKNOWLEDGMENTS

We are grateful to L.B.Litinski for valuable discussions and comments. This
work was supported, in part, by the INTAS-PFBR (project No. 95-0591), by
RSFR (Grant No. 95-02-04239-a), by Russian National program on
Superconductivity (Grant No. 93080) and International Soros Science 
Education Program.

\newpage
\centerline{Figure caption}
\vskip 1cm

Fig.1: (a) The spectra  
$\varepsilon_1({\bf k})$ and
$\varepsilon_2({\bf k})$ calculated for the basis (2),(3),
and the spectrum
$\underline{\varepsilon}_1({\bf k})$ calculated for the basis 
(\ref{oper16}). The spectra are given along symmetrical lines 
$\Gamma -X- M-N-\Gamma$ and $X-N-X$; $\Gamma=(0,0)$; $X=(\pi,0), 
(0,\pi)$, $M=(\pi,\pi)$; $N=(\pi/2,\pi/2)$;

(b) Spectrum equal-energy lines  $\varepsilon_1({\bf k})=const$;

(c) $Z_h({\bf k})$ and $\underline{Z}_h({\bf k})$ --- the number of 
bare holes (the residues of corresponding Green's functions) in the 
quasiparticle excitations for the spectra $\varepsilon_1({\bf k})$ and
$\underline{\varepsilon}_1({\bf k})$; 
$Z_Q({\bf k})$ is the residue of the lowest pole $\varepsilon_1({\bf k})$;
for the Green's function $G_{11,11}(\omega,{\bf 
k})+G_{12,12}(\omega,{\bf k})$, which characterize the "shadow band" 
effect;

(d) the dependence of the number of holes $n$ per unit cell on the 
value of the Fermi surface area $S$ ($S_{BZ}$ -- the area of the first 
Brillouine zone): thick line --- for the spectrum
$\varepsilon_1({\bf k})$; dashed line --- for the spectrum
$\underline{\varepsilon}_1({\bf k})$; solid straight line --- the 
case of the noninteracting particle filling.


\begin{thebibliography}{99}
\bibitem{dagotto}  E.Dagotto, Rev.Mod.Phys. {\bf 66}, 763 (1994).

\bibitem{brenig}  W.Brenig, Physics Reports {\bf 251}, 4 (1995).

\bibitem{emery}  V.J.Emery, Phys.Rev.Lett. {\bf 58}, 2794 (1987);
V.Emery and G.Reiter, Phys.Rev.B {\bf 38}, (1988) 4547.

\bibitem{zh}  F.C.Zhang and T.M.Rice, Phys.Rev. B {\bf 37}, 3759 (1988).

\bibitem{bar}  A.F. Barabanov, L.A. Maksimov and G.V. Uimin, Pisma v
Zh.Exp.Teor.Fis., {\bf 47}, 532 (1988) [JETP Lett. {\bf 47}, 622 (1988)];
Zh.Exp.Teor.Fis. {\bf 96}, 655 (1989) [JETP {\bf 69}, 371 (1989)];
A.F.Barabanov, R.O.Kuzian and L.A.Maksimov, J.Phys.Cond.Matter {\bf 3}, 9129
(1991).

\bibitem{matsukava}
H.Matsukava and H.Fukuyama, J.Phys.Soc.Jap. {\bf 58}, (1989) 2845.

\bibitem{key}
A.F.Barabanov, V.M.Berezovsky, L.A.Maksimov and E.Zhasinas,
Physica C, {\bf 252}, 308 (1995); Zh.Eksp.Teor.Fiz. {\bf 110}, 1480 
(1996) [JETP, {\bf 83}, 819, (1996)].

\bibitem{schrieffer}  J.R.Schrieffer, J.Low.Temp.Phys. {\bf 99}, 397 (1995).

\bibitem{berezovsky}
A.F.Barabanov, V.M.Berezovsky, Phys.Lett.A {\bf 186}, (1994) 175; 
\hskip 6pt Zh.  Eksp. Teor. Fiz. {\bf 106}, (1994) 1156; (JETP {\bf 
79}, (1994) 627).

\bibitem{hayn}
R.Hayn,V. Yushankhai,S.Lovtsov  Phys.Rev.B {\bf 47}, 5253 (1993).

\bibitem {kam}
A.P.Kampf and J.R.Schrieffer,Phis.Rev.B {\bf 42}, 7967 (1990).

\bibitem {mar_hor}
G.Martinez and P.Horsch,Phis.Rev.B {\bf 44}, 317 (1991).

\end{thebibliography}
 
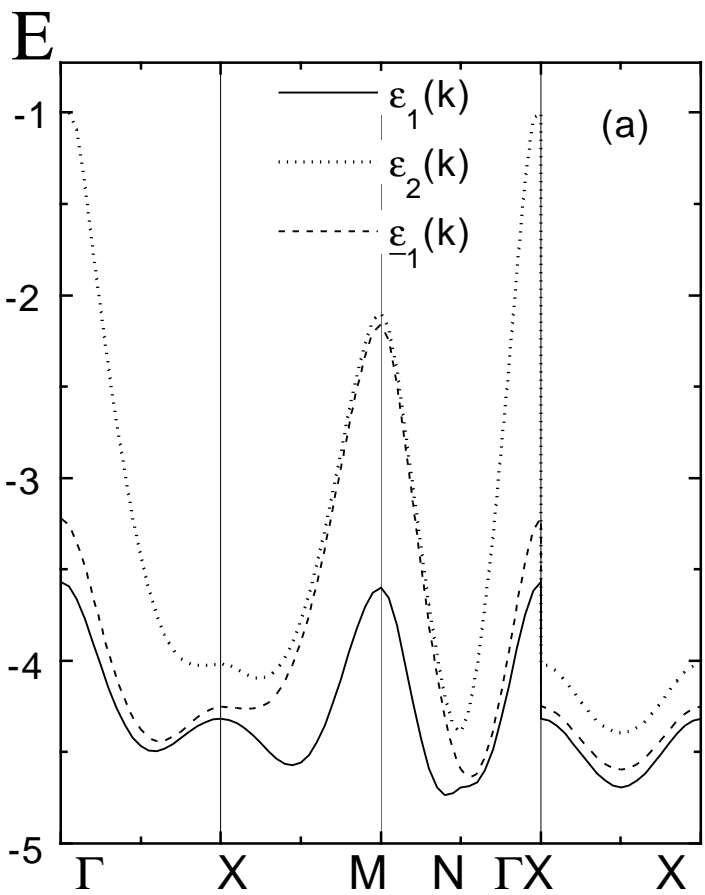
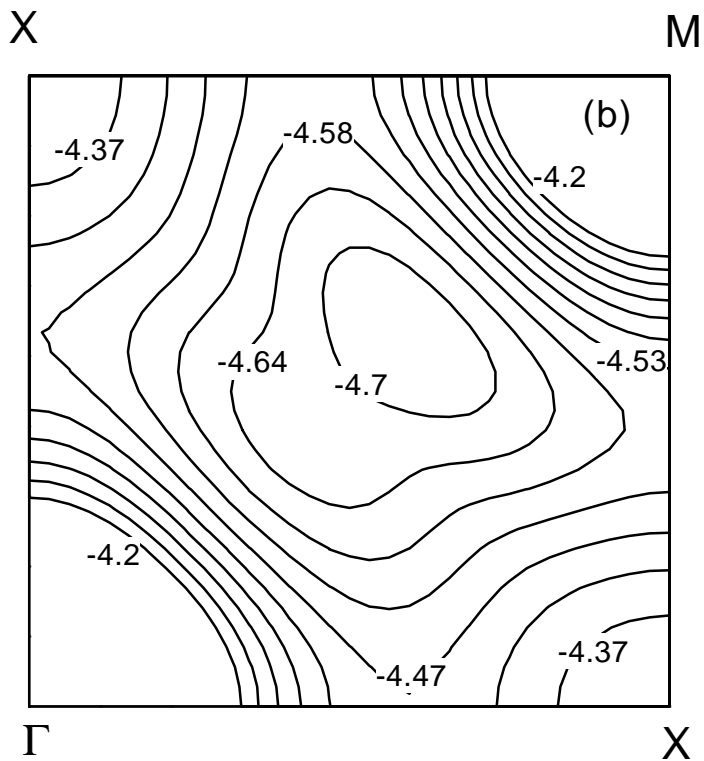
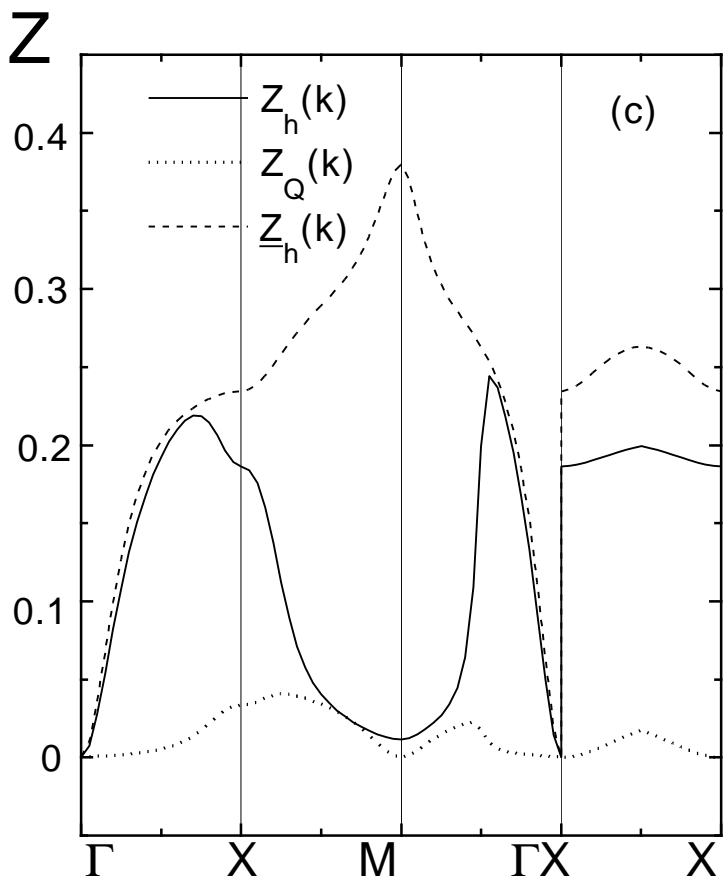
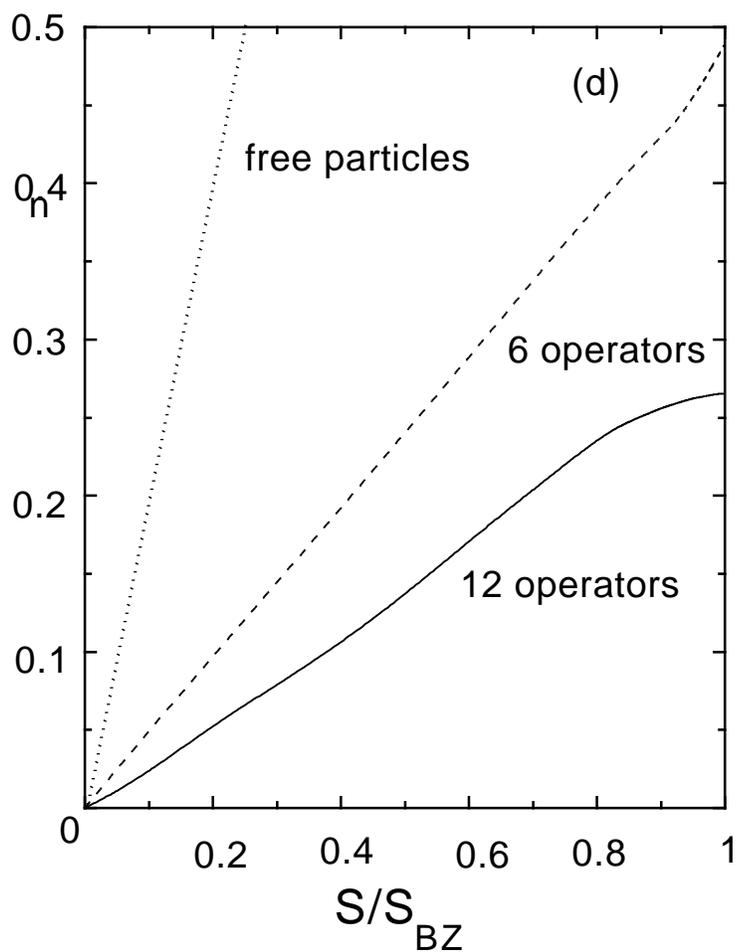
\end{document}